\documentclass[manuscript]{aastex}
\newcommand{\science}{Science}

\shorttitle{WASP-4}
\shortauthors{Petrucci et al.}

\usepackage{graphics}
\usepackage{graphicx}
\begin{document}

\title{NO TRANSIT TIMING VARIATIONS IN WASP-4}

\author{R. Petrucci\altaffilmark{1,4,5},
  E. Jofr\'e\altaffilmark{2,4,5}, M. Schwartz\altaffilmark{1,4},
  V. C\'uneo\altaffilmark{2,4}, C. Mart\'inez\altaffilmark{2},
  M. G\'omez\altaffilmark{2,4}, A. P. Buccino\altaffilmark{1,3,4} and
  P. J. D. Mauas\altaffilmark{1,4}}

\altaffiltext{1}{Instituto de Astronom\'{i}a y F\'{i}sica del Espacio (IAFE), Buenos Aires, Argentina.}
\altaffiltext{2}{Observatorio Astron\'{o}mico de C\'{o}rdoba, C\'{o}rdoba, Argentina.}
\altaffiltext{3}{Departamento de F\'\i sica, FCEN, Universidad de Buenos Aires.}
\altaffiltext{4}{CONICET, Consejo Nacional de Investigaciones Cient\'{i}ficas y T\'{e}cnicas, Argentina.}
\altaffiltext{5}{Visiting Astronomer, Complejo Astron\'{o}mico El Leoncito operated under
agreement between the Consejo Nacional de Investigaciones Cient\'{i}ficas y
T\'{e}cnicas de la Rep\'{u}blica Argentina and the National Universities of La
Plata, C\'{o}rdoba and San Juan.}

\begin{abstract}
We present 6 new transits of the system WASP-4. Together with 28 light
curves published in the literature, we perform an homogeneous study
of its parameters and search for variations in the transit's central
times. The final values agree with those previously
reported, except for a slightly lower inclination. We find no 
significant long-term variations in $i$ 
or $R_{P}/R_{\star}$. The $O-C$ mid-transit times do not show
signs of TTVs greater than 54 seconds.   
\end{abstract}


\section{Introduction}

WASP-4b is one of the exoplanets most studied in the literature. Since
its discovery \citep{wilson08}, many observations of this target
have been made and several authors have determined the physical
properties of the host-star and the exoplanet \citep{gil09,winn09, nik12}. These works
reveal that the system is formed by a G7V star with a close-in hot
Jupiter ($M_{p}=1.28 M_{J}$, $R_{p}=1.39 R_{J}$) in a circular
orbit which transits the star every 1.33 days. WASP-4b is a highly
irradiated planet with a radius larger than the one predicted by
models \citep{for07}. One possibility is that the ongoing
orbital circularization provides the heat needed to inflate the
planet \citep{bee11}.

The transit timing
variations (TTVs) technique has become a very promising
method to estimate the mass of a non-transiting planet when it is not
possible to get radial velocity measurements \citep{hym05}. 
Since the time between transits of a single planet should be
constant, variations in this time can be due to the gravitational
interaction with another planet in the system. If both planets show
transits, it is possible to estimate the radius and mass for each of
them, even without spectroscopic observations. In this way, it is
possible to determine the densities of planets orbiting late stars.
This is one of the key aspects of the TTVs technique. 

Different authors carried out TTVs analysis looking for 
another planetary-mass body in the WASP-4 system
without success. However, most of them employed mid-transit times
fitted with different models and error treatments. As it has
been shown \citep{sou12,nas13} the lack of
homogeneity in the analysis technique can lead to wrong conclusions
about TTVs.  

In this work we present the light curves of 6 new transits
of WASP-4b obtained with telescopes located in Argentina, and
perform an homogeneous study of TTVs, analyzing 34 light curves spanning 6 years of
observations. For all these transits we employed the same
fitting procedure and error treatment to
obtain consistent photometric and physical parameters of the star and
the exoplanet.

In Section \S\ref{sec.obs} we present our observations and data reduction, in
Section \S\ref{sec.proc} we describe the procedure used to fit the light curves and the
parameters derived for the 34 transits. In Section \S\ref{sec.eph} we discuss the
new calculated ephemeris. In Section \S\ref{sec.comp} we compare the
results obtained with the fit provided by the Exoplanet Transit Database and, finally, in Section \S\ref{sec.conc} we present the conclusions.  

\section{Observations and data reduction}\label{sec.obs}

We observed 6 transits of WASP-4b between October 2011 and
July 2013 employing two different telescopes: the Horacio
Ghielmetti Telescope (THG) located at the Complejo Astron\'omico El
Leoncito in San Juan (Argentina), and the 1.54 m telescope located at
the Estaci\'{o}n Astrof\'{i}sica
de Bosque Alegre (EABA, C\'{o}rdoba, Argentina). One of these transits 
was observed with both telescopes simultaneously. In the analysis, we considered
these two measurements as independent. In Table \ref{tbl-1}
we show a log of the observations.

The THG is a remotely-operated 40-cm MEADE - RCX 400, with a focal ratio of
$f/8$. The instrument is currently equipped with an Apogee 
Alta U16M camera with 4098 $\times$ 4098, 9 $\mu$m pixels, resulting in a
scale of 0.57"/pix and a 49'$\times$49' field of view. 
At the EABA, 
we used the 1.54 m telescope in the Newtonian focus, equipped with a 3070 $\times$ 2048, 9 $\mu$m pixels
Apogee Alta U9 camera. This camera provides a 
scale of 0.25"/pix and a  8'$\times$12' field of view. For four transits, we employed the Johnson R filter available at both sites, while for the remaining two we made the observations without filter.

At the beginning of each observing night the computer clock was
automatically synchronized with the GPS. The central times
of the images were expressed in Heliocentric Julian Date based on
Coordinated Universal Time ($HJD_{UTC}$). Whenever possible, we observed 90 minutes
before and after each transit to obtain a large number of out-of-transit (OOT)
 data-points to correct possible trends in the light
curves. We took 10 bias frames, 8 dark frames and
between 15 and 20 dome flat-fields. We averaged all the biases and median-combined
 the bias-corrected darks. Finally, the bias- and
dark-corrected flats were median-combined to generate a master flat in
the corresponding band. All the images were processed using standard IRAF\footnote{IRAF is
distributed by the National Optical Astronomy Observatories, which
are operated by the Association of Universities for Research in
Astronomy, Inc., under cooperative agreement with the National
Science Foundation.} tasks.

To obtain instrumental magnitudes with aperture photometry, we
developed an algorithm called FOTOMCC. This is a quasi-automatic
pipeline developed for the IRAF environment using the “DAOPHOT”
package. Initially, FOTOMCC
employs a reference image, previously selected by the user, to
identify the centroids of the stars in all the images.
The optimal size for
the aperture is chosen through the growth-curves technique \citep{how89}. Specifically, we adopted  the aperture size for which the star
magnitude was stable at the level of 0.001 mag.
The thickness of the sky-subtraction area was set to 5 pixels.
 The magnitude errors were those provided by the
DAOPHOT task. 

To carry out the
differential photometry, for every image we first subtracted from the
magnitude of the science star the one of each star
in the field. Then, we computed the standard deviation of all the
magnitude differences obtained in this way and we selected those stars
which gave the light curves with the lower sigma. 
With the selected stars we built a master star 
whose magnitude and error were the average magnitude and
error of all the chosen stars. The final light curve was built by
the subtraction of the magnitudes of the target and the 
master star. For each photometric data, we estimated the formal error
as the quadrature sum of the errors of the target star and the master star. 

Light curves present smooth trends mainly originated by differential
extinction and/or spectral type differences between the comparison and
the target star. To eliminate these slow variations we fitted
a Legendre polynomial to the OOT data-points and modified its order until the
dispersion of the residuals was minimum. In almost all cases we
used a second-order fit, although in some cases a lower dispersion was found
by fitting a straight line. Finally, for each light curve we removed
the fit from all the data (including transit points) and normalized
the OOT to unity. In Figure \ref{fig1} we present our six
light curves, and the best-fit to the data. Errorbars are also shown.

\subsection{Archival light curves}

To study TTVs and for the parameters determination we also
included all other transits publicly available.
We considered in particular 20 light curves found in the
literature: 1 from
\cite{wilson08}, 2 from \cite{winn09}, 1 from \cite{gil09}, 4 from \cite{soj11} and 12 observed by \cite{nik12}. We did not include the 4 transits from \cite{sou09}, since the authors reported failures in the
computer clock which make the mid-transit times unreliable
 \citep{sou13}. We also included 8 transits observed by
amateurs and published in the Exoplanet Transit 
Database (ETD\footnote{\textsf{http://var2.astro.cz/ETD.}}). 
We only analyzed the complete transits with the
four contact points clearly visible.

\section{Light-curves fitting procedure}\label{sec.proc}

\subsection{Photometric parameters}\label{subsec.photpar}

Based on HARPS high signal-to-noise archival spectra of WASP-4, we derived stellar parameters: effective temperature $T_{eff}$, surface gravity $\log g$,  metallicity $[Fe/H]$ and microturbulence $\xi$, using the FUNDPAR code (Saffe 2011).
The parameters obtained from
the analysis are: $T_{eff}=(5436 \pm 34)$ K, $\log g= (4.28 \pm 0.06) $ cm/s,
$\xi=(0.94 \pm 0.03)$ km/s, $[Fe/H]=(-0.05 \pm 0.04)$ dex (Jofr\'{e} et al. in
preparation). These agree with previously reported values,
except for $\log g$ which is slightly lower (e.g. \citealt{doy13}). 

These stellar parameters were adopted as initial input for the program
JKTLD\footnote{\textsf{http://www.astro.keele.ac.uk/~jkt/codes/jktld.html.}},
which calculates theoretical limb-darkening coefficients
by bilinear interpolation of the effective temperature and surface
gravity using different tabulations. In particular, we employed the
tabulations provided by \cite{van93} and \cite{cla04}. For those
transits observed with no filter we used bolometric limb-darkening coefficients. 	

All the light curves were fitted using the JKTEBOP
code\footnote{\textsf{http://www.astro.keele.ac.uk/~jkt/codes/jktebop.html.}}. This
code models the light curve of a system of two components by
performing numerical integration over the
surface of concentric circles, under the assumption that the projection of each
component is a biaxial ellipsoid. It employs the Levenberg-Marquardt
optimization algorithm to get the best-fitting model. One of the
advantages of JKTEBOP over other fitting models is that it considers
small distortions from sphericity. Since WASP-4b is
a bloated planet, this program can give more realistic parameters from
the observed data.

For each transit, we ran JKTEBOP following the same fitting
procedure:

1) We  assumed as free parameters: the inclination of the orbit ($i$), the sum of
the fractional radii\footnote{$r_{\star}=R_{\star}/a$ and $r_{P}=R_{P}/a$ 
are the ratios of the absolute radii (of the star and the exoplanet respectively) to the semimajor axis.}
($r_{\star} + r_{P}$), the ratio of the fractional
radii ($k=r_{\star}/r_{P}$) and the mid transit time           
($T_{0}$). We fitted every light curve with the linear,         
quadratic, logarithmic and square-root limb-darkening laws. For each
case, we tried with a) both coefficients fixed, b) the linear coefficient fitted
and the nonlinear fixed and c) both coefficients fitted. Finally, we
adopted as the best model for a given transit the one which minimizes the
$\chi^2$ of the fit and gives realistic parameters. 

tbf{ 2) For a few transits, the convergence of some of the adjusted parameters was not achieved in 1). In these cases, assuming the limb-darkening law obtained in the first step, we iterated JKTEBOP taking as initial parameters of each iteration those obtained in
the previous one. This process was repeated until convergence.
     
3) For the solution achieved in 2), we first multiplied the photometric errors 
by the square-root of the reduced chi-squared of the fit to 
get $\chi_{r}^2=1$. Then, we ran the three algorithms available in JKTEBOP: 
Bootstrapping and Monte Carlo simulations and   
Residuals Permutation (RP), which takes red noise into account. For the
first two options we performed 1000 iterations. We conservatively adopted as
the final  errors of the parameters the largest values given by these
algorithms. 

We adopted as the final value for every parameter the median of those
 obtained for every transit (except for $T_0$, see \S\ref{sec.eph}). We
adopted as the final error 
the asymmetric uncertainties $\sigma_{+}$ and $\sigma_{-}$ 
of the selected
distribution, since they
are based on the empirical data and are more realistic than those
derived by a Gaussian distribution of the parameters.

\subsection{Physical parameters}\label{subsec.bozomath}

The physical parameters were determined using standard formulae
\citep{sousolo09} implemented in the JKTABSDIM
code\footnote{\textsf{http://www.astro.keele.ac.uk/~jkt/codes/jktabsdim.html.}}. This
code requires as input the measured quantities: $i$, $r_{\star}$,
$r_{P}$, the orbital period $P$, the velocity amplitudes of the star
 and the exoplanet, $K_{\star}$ and $K_{P}$ respectively, the eccentricity $e$, $T_{eff}$, $[Fe/H]$ and their
errors. For each light curve, we employed the photometric parameters
($i$, $r_{\star}$, $r_{P}$)\footnote{The error considered as input was
the larger between $\sigma_{+}$ and $\sigma_{-}$.} obtained with the
program JKTEBOP, $P$ determined from the
ephemeris, and $T_{eff}$ and $[Fe/H]$ derived using 
HARPS spectra. We used $e=0$ , and the $K_{\star}$ value given by
 \cite{tri10}. 
The procedure was the following: First, assuming $K_{P}=150\,km/s$ we calculated a stellar mass (see Eq. (5) of \cite{sousolo09}). By  
linearly interpolating this stellar mass and the $[Fe/H]$ calculated in \S\ref{subsec.photpar} within tabulated theoretical model, we determined a predicted radius ($R^{(calc)}_{\star}$) and effective
temperature ($T^{(calc)}_{eff}$) for the star. Then, we evaluated the figure of merit:
\begin{equation}
fom = \Bigg[\frac{r^{(obs)}_{\star}-(R^{(calc)}_{\star}/a)}{\sigma(r^{(obs)}_{\star})}\Bigg]^2+\Bigg[\frac{T^{(obs)}_{eff}-T^{(calc)}_{eff})}{\sigma(T^{(obs)}_{eff})}\Bigg]^2 
\end{equation}
We repeated this process until finding the value for $K_{P}$ which minimizes Eq. (1).
In order to avoid any dependence with the stellar-model, we performed this analysis
for 4 different sets of stellar models: $Y^2$ \citep{dem04},
Padova \citep{gir00}, Teramo \citep{pie04} and VRSS \citep{van06}. 
We adopted as the final value for $K_{P}$ the average of
the amplitudes given by each model, and the standard deviation as the error of the velocity.
Finally, the solution for the system was determined using the JKTABSDIM
code.
From this procedure, we also estimated the age of the system considering
series of models bracketing the lifetime of the star in the main
sequence. 

The resulting physical parameters of the star and
the planet obtained for each transit are listed in Table
\ref{tbl-2}. For the exoplanet, the surface gravity was calculated
with:
\begin{equation}
g_{P} = \frac{2\pi}{P}\frac{\sqrt{(1-e^2)} K_{\star}}{r^2_{P} sin(i)}
\end{equation} 
\citep{sou07} and the modified equilibrium temperature as:
 \begin{equation}
T'_{eq} = T_{eff}\sqrt{\frac{R_{\star}}{2a}}
\end{equation}   
\noindent(Southworth 2010). Therefore, both $g_{P}$ and $T^{\prime}_{eq}$ are independent of the stellar             
models. We performed a weighted average of all the measurements to
obtain the final value for each parameter, and the uncertainty
was determined as the standard deviation of the sample. Table \ref{tbl-3} shows
the final values and errors calculated for the photometric and
physical parameters of the star and the exoplanet.  All these 
are in good agreement with previous determinations, except for a
slightly lower inclination. 

The presence of a perturber in the system could produce long-term
variations in these parameters (\citealt{sar99}, \citealt{car10}). Considering that our data comprises 6
years of observations, we studied the long-term behaviour of
$i$ and $R_{P}/R_{\star}$ (Fig \ref{fig2}a and \ref{fig2}b). We found
that these parameters remain constant within the $\pm 1\sigma$ error
of the weighted average, except for the outlier data-point in $i$
corresponding to the epoch 1307, which could have been caused by variable
observing conditions such as the presence of cirrus clouds during that night.

\section{Transit ephemeris and timing}\label{sec.eph}

We transformed the central times of all the observations to $BJD_{TDB}$ (Barycentric 
Julian Date based on Barycentric Dynamical Time) with the \cite{eas10} online converter. For the amateur light curves,
we contacted the observers when extra information was needed. 
For the mid-transit times we adopted the mean values obtained
in Section \S\ref{sec.proc}, and considered
the symmetric errors ($\pm\sigma$) given by the algorithm with the
largest uncertainty.
In most cases, the error 
obtained with the RP method was the largest, indicating the presence of red
noise in the data \citep{pont06}. This implies that there are
correlations between adjacent data points in a light curve, reducing
the number of free parameters. The existence of red noise leads to an
underestimation of the errors in the adjusted parameters which, in
turn, might cause an inaccurate determination of the central time of
the transit. The red noise can be quantified with the factor
$\beta=\sigma_{r}/\sigma_{N}$, defined by \cite{winn08}. Here, 
$\sigma_{r}$ is obtained by averaging the residuals into M bins of N
points and calculating the standard deviation of the binned residuals,
and $\sigma_{N}$ is the expected deviation, calculated by: 
\begin{equation}
\sigma_{N} = \frac{\sigma_{1}}{\sqrt{N}}\sqrt{\frac{M}{M-1}}
\end{equation} 
where $\sigma_{1}$ is the standard deviation of the unbinned
residuals. Considering that the duration of the ingress/egress of the
WASP-4b transits is about 20 minutes, we averaged the residuals in bins 
of between 10 and 30 minutes and calculated the parameter $\beta$ for
each case. Finally, we used the median value as
the red noise factor corresponding to that light curve. In the
absence of red noise, we expect $\beta =1$. For these transits
$\beta$ ranges from 0.58 to 2.36.                                         

The whole sample of
mid-transit times presents 2 big outliers corresponding to the epochs       
298 and 1085. The first transit was obtained from the ETD.
In the latter case, we believe there was a failure in
the computer clock.  We did not considered these points for further
analysis. Therefore, we determined the ephemeris in three different ways: a) considering all
the 32 remaining transits, b) excluding the incomplete transit (indicated in Fig \ref{fig1} as 2013-06-06 and observed at EABA), and c) only considering
those transits with $\beta \le 1.6$ (30 points). In the three cases
we fitted the data through weighted least-squares to obtain the best
period and the minimum reference time. We re-scaled the uncertainties
multiplying them by $\sqrt{\chi^2_{r}}$.  The final values and errors for $P$ and $T_{0}$ obtained from
different sets are:
\begin{eqnarray*}
a)\,P=1.33823251(31)\,\textrm{days},\,T_{0}=2454697.797973(76)\,BJD_{TDB}\\ 
b)\,P=1.33823251(32)\,\textrm{days},\,T_{0}=2454697.797973(77)\,BJD_{TDB}\\
c)\,P=1.33823227(32)\,\textrm{days},\,T_{0}=2454697.797973(77)\,BJD_{TDB} \\
\end{eqnarray*}
\noindent {Since there are no differences in $T_{0}$, the inclusion of partial
transits, or those obtained with large red noise, does not affect the
calculation. We adopted the ephemeris given by the sample a)
including all the transits.

With the new ephemeris, we calculated the $O-C$ mid-transit times, which
are shown in Figure \ref{fig2}c. Except for the already mentioned outliers, all
differences are within the $\pm 1\sigma$
error. The RMS of the data is 54 seconds. We ran a Lomb-Scargle
periodogram \citep{hor86} to the data, excluding the 2 big
outliers, and no significant peak was found.

\section{Comparison between JKTEBOP and the fitting model in the ETD}\label{sec.comp}

For the light curves taken from the ETD, we 
compared the mid-transit times obtained with JKTEBOP and those given by the ETD,
which provides an
automatic fit, modeling the photometric data with the 
function (Poddany et al. 2010): 
\begin{equation}
m(t_{i}) = A-2.5\log F(z[t_{i},t_{0},D,b], p, c_{1}) + B(t_{i}-t_{mean})+ C(t_{i}-t_{mean})^{2}
\label{eq.mod}
\end{equation}
where $m(t_{i})$ are the relative magnitudes taken at the times
$t_{i}$, $N$ is the number of data-points,
$t_{mean} =t_{i}/N$ is the mean time of the observations,
$z$ is the projected relative-separation of the planet from 
the star, $p$ is the ratio of the planet to star radii and $F(z,p,
c_{1})$ is the $occultsmall$ routine of \cite{man02}, giving 
the relative flux of the star as the planet transits.    
This model assumes a linear limb-darkening law with the 
coefficient $c_{1}$ fixed at an arbitrary value of $0.5$. 
The user has the possibility to fit or maintain fixed 
the mid-transit time, the duration and the depth parameters. 
The coefficients of Eq. (\ref{eq.mod}) are calculated using the
Levenberg-Marquardt non-linear least-squares algorithm 
from \cite{press92}. The optimal parameters are determined by
iterating the procedure until the difference between two
successive values of $\Delta\chi^2$ is negligible.

We fitted the three parameters simultaneously and converted the
resulting $HJD_{UTC}$ mid-transit times to $BJD_{TDB}$. In Figure \ref{fig2}d we
show the differences between the central times determined in both
ways. The errorbars are those derived using the ETD model. 
The differences are as large as 1.5                      
minutes. We believe these disagreements are due to the very simple
limb-darkening law assumed in the ETD fit. In any case, these differences point out to 
the need to derive the mid-transit times with an homogeneous method,
when searching for TTVs.

\section{Summary and conclusions}\label{sec.conc}                   

In this work we present 6 new observations of transits of WASP-4b, observed between
2011 and 2013. Using these observations together with another 28
transits previously reported (including 8 observed by amateurs), we
performed an homogeneous study of the system taking into 
account the realistic possibility of distortions in its components.
The physical parameters of the star and the exoplanet are consistent 
with previous determinations, except for the inclination which is slightly
lower, probably due to the fitting procedure. 

In addition, we analyzed the long-term behaviour of different parameters.
Except for one outlier in $i$, and two 
for the $O-C$
mid-transit times, all these parameters remain stable within the
$\pm 1\sigma$ error of the weighted averages. The RMS of the
mid-transit times is 54 seconds. Therefore, we 
confirm previous
results, and found that the system does not show significant TTVs
attributable to the presence of a perturber, a conclusion 
we expanded with two more years of observations, to a baseline of 6
years. The lack of temporal variations in the
rest of the parameters supports this conclusion. 

Finally, we report differences as large as 1.5 minutes between the 
mid-transit times modeled by the fitting programs provided  by the
ETD and JKTEBOP. Therefore, we believe that the central times provided
by the ETD should be used with caution in TTV studies. 

\acknowledgments

We are grateful to Pablo Perna and the CASLEO staff for technical support, and to CONICET for funding  this research. We thank the anonymous referee for their useful comments, and
Phil Evans, Ivan Curtis, and T. G. Tan for kindly providing us
with information about their observations.

\bibliographystyle{apj}

\begin{thebibliography}{}
\bibitem[Beerer et al.(2011)]{bee11} Beerer, I. M., Knutson, H. A., Burrows, A., et al., 2011, \apj, 727, 23
\bibitem[Carter \& Winn(2010)]{car10}  Carter J. A., \& Winn J. N., 2010, \apj, 716, 850
\bibitem[Claret(2004)]{cla04}  Claret, A., 2004, A$\&$A, 428, 1001
\bibitem[Demarque et al.(2004)]{dem04}  Demarque, P., Woo, J. H., Kim, Y. C., Yi, S. K., 2004, \apjs, 155, 667
\bibitem[Doyle et al.(2013)]{doy13} Doyle, A. P., Smalley, B., Maxted, P. F. L., Anderson, D. R., et al., 2013, \mnras, 428, 3164
\bibitem[Eastman et al.(2010)]{eas10} Eastman, J., Siverd, R., Gaudi, B. S., 2010, \pasp, 122, 935
\bibitem[Fortney et al.(2007)]{for07}  Fortney, J. J., Marley, M. S., Barnes, J. W., 2007, \apj, 668, 1267
\bibitem[Gillon et al.(2009)]{gil09}  Gillon, M., Smalley, B., Hebb, L. et al., 2009, A$\&$A, 496, 259
\bibitem[Girardi et al.(2000)]{gir00}  Girardi, L., Bressan, A., Bertelli, G., Chiosi, C., 2000, A$\&$AS, 141, 371
\bibitem[Holman \& Murray(2005)]{hym05} Holman, M. J., \& Murray, 2005, \science, 307, 1288
\bibitem[Horne \& Baliunas(1986)]{hor86} Horne, J. H., \& Baliunas, S. L., 1986, \apj, 302, 757
\bibitem[Howell(1989)]{how89} Howell, S. B., 1989, \pasp, 101, 616
\bibitem[Mandel \& Agol(2002)]{man02} Mandel, K., \& Agol, E., 2002, \apjl, 580, 171
\bibitem[Nascimbeni et al.(2013)]{nas13} Nascimbeni, V., Cunial, A., Murabito, S., et al.,  2013,  A$\&$A, 549, 30
\bibitem[Nikolov et al.(2012)]{nik12} Nikolov N., Henning T., Koppenhoefer J., et al., 2012,  A$\&$A, 539, A159
\bibitem[Pietrinferni et al.(2004)]{pie04} Pietrinferni, A., Cassisi, S., Salaris, M., Castelli, F., 2004, \apj, 612, 168
\bibitem[Poddany et al.(2010)]{pod10} Poddany, S., Brat, L., Pejcha, O., 2010, NewAstronomy, 15, 297
\bibitem[Pont et al.(2006)]{pont06} Pont, F., Zucker S., Queloz, D., 2006, MNRAS, 373, 231
\bibitem[Press et al.(1992)]{press92} Press, W. H., Teukolsky, S. A., Vetterling, W. T., Flannery, B. P., 1992. Numerical Recipes in C. The Art of Scientific Computing. University Press, Cambridge.
\bibitem[Saffe(2011)]{saf11} Saffe, C., 2011, RMxAA, 47, 3
\bibitem[Sanchis-Ojeda et al.(2011)]{soj11} Sanchis-Ojeda R., Winn J. N., Holman M. J., et al., 2011, \apj, 733, 127
\bibitem[Sartoretti \& Schneider(1999)]{sar99} Sartoretti P., \& Schneider J., 1999, A$\&$AS, 134, 553
\bibitem[Southworth (2009)]{sousolo09} Southworth, J., 2009, \mnras, 394, 272
\bibitem[Southworth(2010)]{sou10} Southworth, J., 2010, \mnras, 408, 1689 
\bibitem[Southworth et al.(2007)]{sou07} Southworth, J., Wheatley, P. J., \& Sams, G., 2007, \mnras, 379, 11
\bibitem[Southworth et al.(2009)]{sou09} Southworth, J., Hinse, T. C., Burgdorf,  M. J., et al., 2009, \mnras, 399, 287 
\bibitem[Southworth et al.(2012)]{sou12} Southworth, J., Bruni, I., Mancini,  L., Gregorio, J., 2012, \mnras, 420, 2580
\bibitem[Southworth et al.(2013)]{sou13} Southworth, J., Mancini, L., Browne, P., et al., 2013, accepted for publication in \mnras 
\bibitem[Triaud et al.(2010)]{tri10} Triaud, A. H. M. J., et al., 2010, A$\&$A, 524, 25
\bibitem[Van Hamme(1993)]{van93} Van Hamme, W., 1993, \aj, 106, 2096 	
\bibitem[VandenBerg et al.(2006)]{van06} VandenBerg, D. A., Bergbusch, P. A., Dowler, P. D., 2006, \apjs, 162, 375
\bibitem[Wilson et al.(2008)]{wilson08} Wilson, D. M., Gillon, M., Hellier, C., et al., 2008, \apj, 675, 113
\bibitem[Winn et al.(2008)]{winn08} Winn, J. N., Holman, M. J., Torres, G., et al., 2008, \apj, 683, 1076
\bibitem[Winn et al.(2009)]{winn09} Winn, J. N., Holman, M. J., Carter, J. A., et al., 2009, \aj, 137, 3826 
\end{thebibliography}

\clearpage

\begin{figure}
\epsscale{1.00}
\plotone{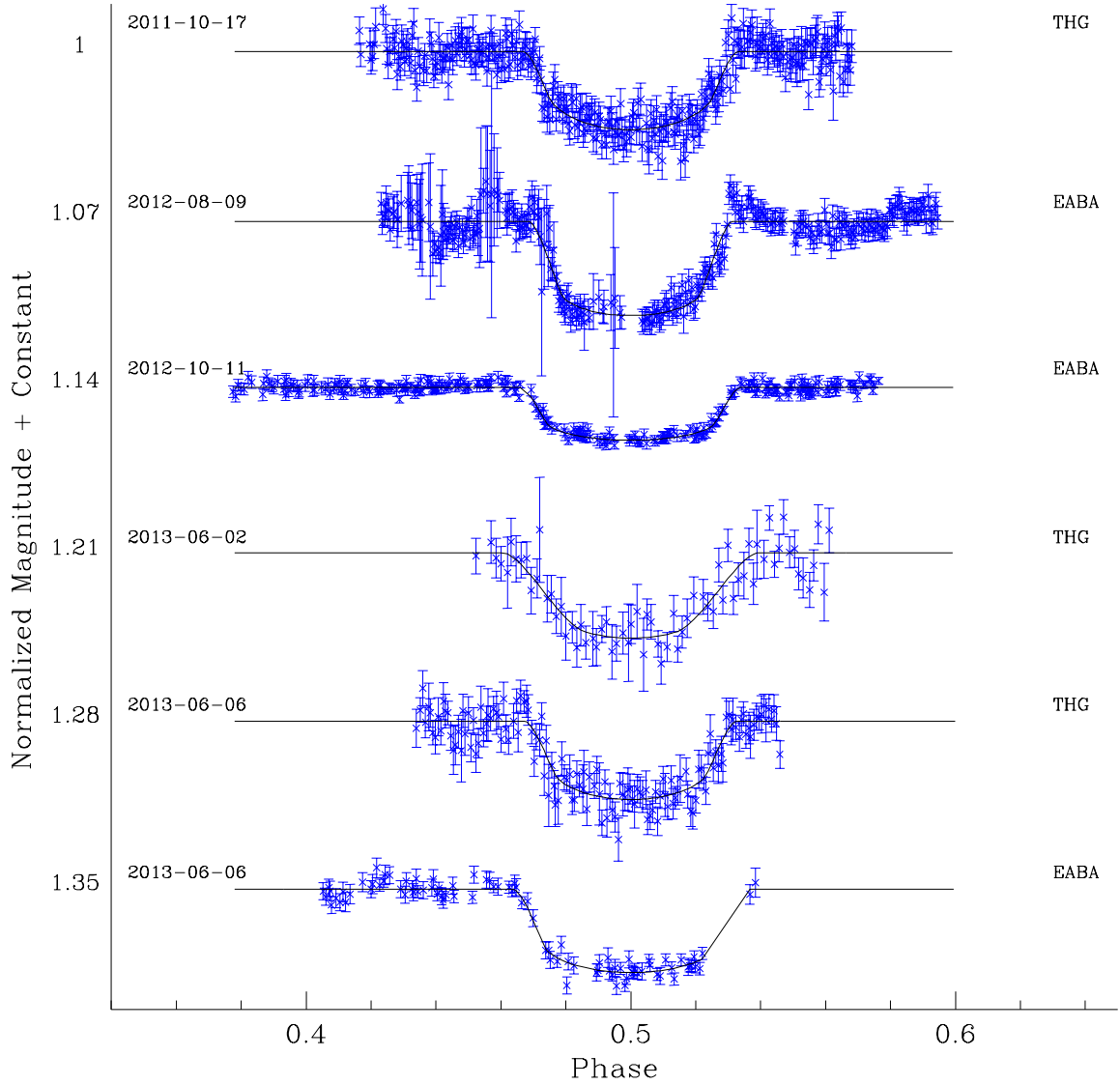}
\caption{Light-curves presented in this work. The photometric
  observations and their error-bars are in blue. Black solid lines
  represent the best-fit to the data. For each transit, the date and the telescope are indicated.\label{fig1}} 
\end{figure}

\begin{figure}
\epsscale{.80}
\plotone{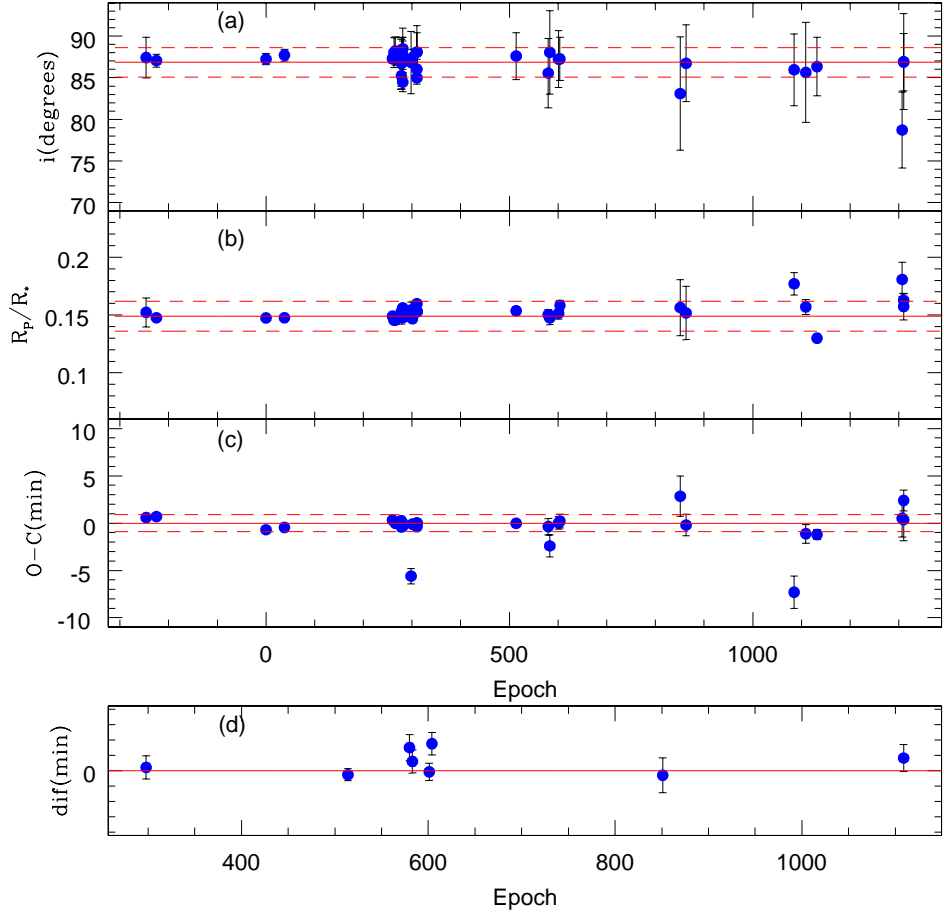}
\caption{Figures a, b and c: $i$, $R_{P}/R_{\star}$ and $O-C$ as a
  function of the transit epoch. Blue points are the values obtained
  for the 34 light curves. The solid and dashed horizontal red lines
  represent the weighted average and its $\pm1 \sigma$ errors,
  respectively. Figure d: Differences between the central transit
  times obtained with the model used in the ETD and the ones obtained in this work.}.\label{fig2}
\end{figure}

\clearpage

\begin{deluxetable}{cccccccc}
\tabletypesize{\scriptsize}
\tablecaption{Log of our observations \label{tbl-1}}
\tablewidth{0pt}
\tablehead{
\colhead{Date} & \colhead{Telescope} & \colhead{Camera} & \colhead{Filter} & \colhead{Bin-size} &
\colhead{Exposure-Time (s)} & \colhead{$N_{obs}$\tablenotemark{a}} & \colhead{$\sigma$\tablenotemark{b}(mag)} 
}
\startdata
Oct 17 2011 & THG & U16M & no filter & 1x1  & 25 & 315 & 0.0059 \\
Aug 9 2012 & EABA & U9 & no filter & 1x1   & 40-50 & 306 & 0.0069   \\
Oct 11 2012 & EABA & U9 & R & 2x2   & 25  & 272 & 0.0019 \\
Jun 2 2013 & THG & U16M & R & 2x2   &  180  & 68 & 0.0090\\
Jun 6 2013 & THG & U16M & R & 2x2   & 45  & 90 & 0.0063\\
Jun 6 2013 & EABA & U9 & R & 2x2   & 90 & 141 & 0.0035 \\
\enddata
\tablenotetext{a}{ Number of data-points.}
\tablenotetext{b}{ Standard deviation of the out-of-transit data-points.} 
\end{deluxetable}

\clearpage

\begin{deluxetable}{ccccccccccccc}
\tabletypesize{\scriptsize}
\rotate
\setlength{\tabcolsep}{0.025in}
\tablewidth{0pt}
\tablecaption{Physical properties of the star and the exoplanet derived in this work for the 34 light curves \label{tbl-2}}
\tablehead{
\colhead{Epoch}      & \colhead{Date} & \colhead{$i$($\degr$)} &
\colhead{$M_{P}$ ($M_{Jup}$)}          & \colhead{$R_{P}$ ($R_{Jup}$)}  &
\colhead{$g_{P}$ [m/s$^{2}$]}          & \colhead{$T^{\prime}_{eq}$ (K)}      &
\colhead{$M_{\star}$ ($M_{\odot}$)}          & \colhead{$R_{\star}$ ($R_{\odot}$)}  &
\colhead{$\log g_{\star}$ [cm/s]}          & \colhead{Age (Gyr)}    &
\colhead{a (AU)}  & \colhead{Author}}
\startdata
-246 & 2007-09-25 &  87.42 $\pm$ 2.46 &	1.22	$\pm$	0.032	&	1.349	$\pm$	0.088	&	16.6	$\pm$	2.8	&	1657	$\pm$	56.5	&	0.894	$\pm$	0.031	&	0.915	$\pm$	0.052	&	4.467	$\pm$	0.048	&	5.1	$\pm$	3	&	0.02289	$\pm$	0.00026	& 1\\
-225	& 2007-10-07 &   87.05 $\pm$ 0.76 &	1.218	$\pm$	0.03	&	1.31	$\pm$	0.028	&	17.5	$\pm$	1.4	&	1657	$\pm$	22.5	&	0.891	$\pm$	0.029	&	0.914	$\pm$	0.015	&	4.466	$\pm$	0.012	&	5.5	$\pm$	3	&	0.02287	$\pm$	0.00025	& 2\\
0	&  2008-08-19 &	87.23 $\pm$ 0.68 & 1.215	$\pm$	0.031	&	1.323	$\pm$	0.028	&	17.2	$\pm$	1.4	&	1666	$\pm$	23.5	&	0.888	$\pm$	0.03	&	0.923	$\pm$	0.015	&	4.456	$\pm$	0.012	&	6.8	$\pm$	3	&	0.02284	$\pm$	0.00026	& 3 \\
38	& 2008-10-09 & 87.72 $\pm$ 0.72 &	1.226	$\pm$	0.034	&	1.314	$\pm$	0.025	&	17.5	$\pm$	1.3	&	1654	$\pm$	22.9	&	0.9	$\pm$	0.033	&	0.914	$\pm$	0.014	&	4.471	$\pm$	0.01	&	5.1	$\pm$	3.1	&	0.02294	$\pm$	0.00028	& 3 \\
260	& 2009-08-02 &	87.29 $\pm$ 0.59 & 1.217	$\pm$	0.028	&	1.329	$\pm$	0.023	&	17	$\pm$	1.2	&	1660	$\pm$	19	&	0.891	$\pm$	0.026	&	0.918	$\pm$	0.012	&	4.462	$\pm$	0.009	&	5.7	$\pm$	2.7	&	0.02286	$\pm$	0.00022	& 4 \\
263	& 2009-08-06 &	88.03 $\pm$ 1.80 & 1.218	$\pm$	0.03	&	1.291	$\pm$	0.027	&	18.1	$\pm$	1.4	&	1656	$\pm$	21.1		&	0.892	$\pm$	0.028	&	0.914	$\pm$	0.014	&	4.467	$\pm$	0.01	&	5.3	$\pm$	2.8	&	0.02287	$\pm$	0.00024	& 4 \\
266	& 2009-08-10 & 88.21 $\pm$ 1.70 &	1.217	$\pm$	0.031	&	1.291	$\pm$	0.037	&	18.1	$\pm$	1.7	&	1656	$\pm$	24.4	&	0.891	$\pm$	0.03	&	0.913	$\pm$	0.016	&	4.467	$\pm$	0.013	&	5.4	$\pm$	3	&	0.02286	$\pm$	0.00026	& 4 \\
278	& 2009-08-26 & 86.72 $\pm$ 3.07 &	1.206	$\pm$	0.035	&	1.355	$\pm$	0.111	&	16.2	$\pm$	3.3	&	1680	$\pm$	52.8	&	0.877	$\pm$	0.032	&	0.934	$\pm$	0.047	&	4.44	$\pm$	0.043	&	8.8	$\pm$	2.5	&	0.02274	$\pm$	0.00028	& 5 \\
278	& 2009-08-26 &	87.63 $\pm$ 2.23 & 1.219	$\pm$	0.031	&	1.325	$\pm$	0.083	&	17.2	$\pm$	2.8	&	1664	$\pm$	43.5	&	0.893	$\pm$	0.03	&	0.922	$\pm$	0.038	&	4.46	$\pm$	0.035	&	5.9	$\pm$	2.6	&	0.02288	$\pm$	0.00026	& 5\\
278	& 2009-08-26 &	85.23 $\pm$ 1.60 & 1.195	$\pm$	0.035	&	1.447	$\pm$	0.087	&	14.1	$\pm$	2.3	&	1717	$\pm$	51		&	0.863	$\pm$	0.033	&	0.971	$\pm$	0.045	&	4.399	$\pm$	0.039	&	11.2	$\pm$	3.3	&	0.02262	$\pm$	0.00029	& 5 \\
278	& 2009-08-26 & 86.81 $\pm$ 2.81 &	1.208	$\pm$	0.03	&	1.375	$\pm$	0.068	&	15.8	$\pm$	2.1	&	1677	$\pm$	41.8	&	0.879	$\pm$	0.028	&	0.932	$\pm$	0.037	&	4.443	$\pm$	0.033	&	8.5	$\pm$	2	&	0.02276	$\pm$	0.00024	& 5\\
281	& 2009-08-30 & 88.49 $\pm$ 2.47 &	1.231	$\pm$	0.035	&	1.307	$\pm$	0.107	&	17.8	$\pm$	3.6	&	1654	$\pm$	54.6	&	0.906	$\pm$	0.033	&	0.916	$\pm$	0.049	&	4.471	$\pm$	0.046	&	4.1	$\pm$	3	&	0.02299	$\pm$	0.00028	& 5 \\
281	& 2009-08-30 &	84.47 $\pm$ 1.15 & 1.195	$\pm$	0.038	&	1.461	$\pm$	0.069	&	13.8	$\pm$	1.9	&	1711	$\pm$	48	&		0.862	$\pm$	0.037	&	0.964	$\pm$	0.04	&	4.406	$\pm$	0.035	&	11.3	$\pm$	3.7	&	0.02261	$\pm$	0.00032	& 5\\
281	& 2009-08-30 &	87.70 $\pm$ 2.14 & 1.223	$\pm$	0.037	&	1.332	$\pm$	0.093	&	17	$\pm$	3.1	&	1655	$\pm$	60.1	&		0.897	$\pm$	0.036	&	0.913	$\pm$	0.054	&	4.47	$\pm$	0.051	&	4.6	$\pm$	3.4	&	0.02291	$\pm$	0.00031	& 5 \\
281	& 2009-08-30 & 86.89 $\pm$ 2.53 &	1.217	$\pm$	0.036	&	1.367	$\pm$	0.069	&	16.1	$\pm$	2.3	&	1666	$\pm$	43.4		&	0.89	$\pm$	0.034	&	0.923	$\pm$	0.036	&	4.457	$\pm$	0.033	&	6.8	$\pm$	3.1	&	0.02285	$\pm$	0.00029	& 5\\
298	&	2009-09-21 & 86.83 $\pm$ 3.75 & 1.206	$\pm$	0.035	&	1.402	$\pm$	0.248	&	15.1	$\pm$	6	&	1687	$\pm$	136	&		0.877	$\pm$	0.032	&	0.942	$\pm$	0.14	&	4.433	$\pm$	0.13	&	8.8	$\pm$	2.6	&	0.02274	$\pm$	0.00028	& 6 \\
301	&	2009-09-26 & 87.42 $\pm$ 1.00 & 1.218	$\pm$	0.03	&	1.31	$\pm$	0.032	&	17.5	$\pm$	1.5	&	1661	$\pm$	25.9		&	0.892	$\pm$	0.029	&	0.919	$\pm$	0.019	&	4.462	$\pm$	0.016	&	6.2	$\pm$	2.6	&	0.02287	$\pm$	0.00025	& 4 \\
310	& 2009-10-08 &	88.06 $\pm$ 2.34 & 1.234	$\pm$	0.031	&	1.347	$\pm$	0.107	&	16.8	$\pm$	3.3	&	1643	$\pm$	52.8		&	0.909	$\pm$	0.029	&	0.904	$\pm$	0.048	&	4.484	$\pm$	0.046	&	3.7	$\pm$	2.4	&	0.02302	$\pm$	0.00025	& 5\\
310	& 2009-10-08 & 84.99 $\pm$ 0.78 & 1.202	$\pm$	0.03	&	1.484	$\pm$	0.05	&	13.5	$\pm$	1.4	&	1699	$\pm$	37.8		&	0.869	$\pm$	0.029	&	0.953	$\pm$	0.032	&	4.419	$\pm$	0.028	&	10.1	$\pm$	2.6	&	0.02268	$\pm$	0.00025	& 5 \\
310	& 2009-10-08 &	86.02 $\pm$ 1.22 & 1.213	$\pm$	0.029	&	1.382	$\pm$	0.068	&	15.7	$\pm$	2.1	&	1673	$\pm$	44.5		&	0.884	$\pm$	0.027	&	0.929	$\pm$	0.04	&	4.448	$\pm$	0.037	&	7.4	$\pm$	2.3	&	0.0228	$\pm$	0.00023	& 5 \\
310	& 2009-10-08 &	88.09 $\pm$ 3.16 & 1.234	$\pm$	0.031	&	1.333	$\pm$	0.14	&	17.2	$\pm$	4.2	&	1641	$\pm$	74.2		&	0.909	$\pm$	0.029	&	0.903	$\pm$	0.072	&	4.486	$\pm$	0.069	&	3.7	$\pm$	2.3	&	0.02302	$\pm$	0.00025	& 5\\
514	& 2010-07-07 &	87.60 $\pm$ 2.83 & 1.226	$\pm$	0.037	&	1.362	$\pm$	0.159	&	16.3	$\pm$	4.5	&	1653	$\pm$	84.3		&	0.9	$\pm$	0.036	&	0.912	$\pm$	0.081	&	4.472	$\pm$	0.077	&	4.5	$\pm$	3.5	&	0.02294	$\pm$	0.00031	& 7 \\
580	& 2010-10-04 &	85.55 $\pm$ 4.17 & 1.212	$\pm$	0.036	&	1.353	$\pm$	0.201	&	16.4	$\pm$	5.5	&	1669	$\pm$	95.8		&	0.882	$\pm$	0.031	&	0.924	$\pm$	0.095	&	4.452	$\pm$	0.089	&	7.7	$\pm$	2	&	0.02279	$\pm$	0.00027	& 8 \\
583	& 2010-10-07 &	88.05 $\pm$ 5.01 & 1.232	$\pm$	0.035	&	1.293	$\pm$	0.299	&	18.2	$\pm$	9.1	&	1644	$\pm$	153.8		&	0.907	$\pm$	0.033	&	0.905	$\pm$	0.158	&	4.482	$\pm$	0.153	&	4	$\pm$	2.9	&	0.023	$\pm$	0.00028	& 8\\
601	& 2010-11-01 & 87.24 $\pm$ 3.42 &	1.231	$\pm$	0.031	&	1.336	$\pm$	0.202	&	17	$\pm$	5.8	&	1651	$\pm$	143.6	&		0.905	$\pm$	0.029	&	0.912	$\pm$	0.149	&	4.475	$\pm$	0.143	&	3.6	$\pm$	2.3	&	0.02298	$\pm$	0.00024	& 9 \\
604	& 2010-11-05 &	87.26 $\pm$ 2.60 & 1.233	$\pm$	0.03	&	1.38	$\pm$	0.155	&	16	$\pm$	4.2	&	1634	$\pm$	83.6	&		0.908	$\pm$	0.028	&	0.895	$\pm$	0.082	&	4.493	$\pm$	0.08	&	3.6	$\pm$	2.3	&	0.023	$\pm$	0.00024	& 8\\
851	& 2011-10-01 &	83.10 $\pm$ 6.80 & 1.188	$\pm$	0.06	&	1.553	$\pm$	0.796	&	12.2	$\pm$	13.2	&	1779	$\pm$	265.8		&	0.85	$\pm$	0.045	&	1.037	$\pm$	0.292	&	4.336	$\pm$	0.251	&	14.6	$\pm$	1.6	&	0.02251	$\pm$	0.0004	& 6 \\
863	& 2011-10-17 &	86.74 $\pm$ 4.62 & 1.223	$\pm$	0.029	&	1.355	$\pm$	0.235	&	16.5	$\pm$	6.3	&	1672	$\pm$	171.4	&		0.895	$\pm$	0.026	&	0.932	$\pm$	0.182	&	4.451	$\pm$	0.172	&	5.2	$\pm$	1.7	&	0.0229	$\pm$	0.00022	& 10 \\
1085	& 2012-08-09 &	85.95 $\pm$ 4.30 & 1.235	$\pm$	0.036	&	1.51	$\pm$	0.256	&	13.4	$\pm$	5.1	&	1628	$\pm$	156.1		&	0.908	$\pm$	0.031	&	0.888	$\pm$	0.16	&	4.499	$\pm$	0.158	&	3.6	$\pm$	2.2	&	0.02301	$\pm$	0.00026	& 11 \\
1109	& 2012-09-10 &	85.64 $\pm$ 6.01 & 1.202	$\pm$	0.045	&	1.457	$\pm$	0.333	&	14	$\pm$	7.1	&	1698	$\pm$	213		&	0.871	$\pm$	0.039	&	0.953	$\pm$	0.225	&	4.42	$\pm$	0.209	&	9.6	$\pm$	3.1	&	0.02269	$\pm$	0.00033	& 12 \\
1132	& 2012-10-11 &	86.34 $\pm$ 3.50 & 1.204	$\pm$	0.036	&	1.202	$\pm$	0.101	&	20.6	$\pm$	4.3	&	1693	$\pm$	67.3		&	0.874	$\pm$	0.034	&	0.949	$\pm$	0.063	&	4.426	$\pm$	0.057	&	9.2	$\pm$	2.6	&	0.02272	$\pm$	0.00029	& 11 \\
1307	& 2013-06-02 &	78.72 $\pm$ 4.56 & 1.241	$\pm$	0.111	&	2.252	$\pm$	0.477	&	6	$\pm$	3.2	&	1947	$\pm$	222.6		&	0.892	$\pm$	0.108	&	1.262	$\pm$	0.238	&	4.186	$\pm$	0.163	&	13	$\pm$	6.3	&	0.02287	$\pm$	0.00092	& 10\\
1310	& 2013-06-06 &	86.93 $\pm$ 5.76 & 1.22	$\pm$	0.035	&	1.34	$\pm$	0.359	&	16.8	$\pm$	9.7	&	1661	$\pm$	155.6		&	0.894	$\pm$	0.029	&	0.92	$\pm$	0.162	&	4.462	$\pm$	0.155	&	5	$\pm$	1.8	&	0.02289	$\pm$	0.00025	& 10 \\
1310	& 2013-06-06 &	86.89 $\pm$ 3.44 & 1.196	$\pm$	0.032	&	1.53	$\pm$	0.271	&	12.6	$\pm$	5	&	1712	$\pm$	126.3		&	0.867	$\pm$	0.03	&	0.966	$\pm$	0.132	&	4.406	$\pm$	0.119	&	10.8	$\pm$	2.4	&	0.02266	$\pm$	0.00026	& 11\\
\enddata 
\tablecomments{(1) Wilson et al. (2008); (2) Gillon et al. (2009); (3) Winn et al. (2009); (4) Sanchis-Ojeda et al. (2011); (5) Nikolov et al. (2012); (6) Tifner (ETD); (7) Sauer (ETD); (8) Curtis (ETD); (9) TgTan (ETD); (10) This work (THG); (11) This work (EABA); (12) Evans (ETD).}
\end{deluxetable}

\clearpage

\begin{deluxetable}{ccc}
\tablecaption{Final parameters of the WASP-4 system derived in this work\label{tbl-3}}
\tablewidth{0pt}
\tablehead{
\colhead{Parameter} & \colhead{Value} & \colhead{Error} 
}
\startdata
Period $P$(days) & 1.33823251 & 0.00000031   \\
Minimum reference Time $T_{0}$($BJD_{TDB}$) & 2454697.797973 & 0.000076    \\
Inclination $i$($\degr$) & 86.85  & 1.76   \\
Stellar Radius $R_{\star}$($R_{\sun}$) & 0.92 & 0.06   \\ 
Stellar Mass $M_{\star}$($M_{\sun}$) & 0.89  &  0.01   \\
Stellar gravity $\log g_{\star}$($cm/s$) & 4.461 &  0.054 \\
Semimajor axis $a$($UA$) & 0.0228  & 0.00013   \\
Age (Gyr) & 7.0 & 2.9   \\  
Stellar effective temperature $T_{eff}$(K) & 5436 & 34  \\
Metallicity $[Fe/H]$(dex) & -0.05 & 0.04   \\
Planet Radius $R_{P}$($R_{Jup}$) & 1.33  &  0.16  \\
Planet Mass $M_{P}$($M_{Jup}$) & 1.216 &   0.013  \\
Planet surface gravity $g_{P}$($m/s^{2}$) & 16.41 &  2.49 \\
Planet equilibrium temperature $T'_{eq}$(K) & 1664 & 54   \\
\enddata
\end{deluxetable}

\end{document}